\documentclass[%
 reprint,
 amsmath,amssymb,
 aps,
]{revtex4-2}

\usepackage{graphicx}
\usepackage{dcolumn}
\usepackage{bm}

\usepackage[T1]{fontenc}
\usepackage{makecell}
\usepackage{multirow}
\usepackage{booktabs}
\usepackage{microtype}

\newtheorem{definition}{Definition}
\newcommand{\ROMAN}{\textls*[-100]}
\newcommand{\tabitem}{~~\llap{\textbullet}~~}

\begin{document}

\preprint{APS/123-QED}

\title{Conformity to continuous and discrete ordinal traits}

\author{Elisa Heinrich Mora,$^{\text{a,b},1}$ Kaleda K. Denton,$^{\text{a,b},1}$ Michael E. Palmer,$^\text{a}$ Marcus W. Feldman$^{\text{a},2}$ \\ \vspace{0.5cm} $^1$Equal contributions \\ $^2$To whom correspondence should be addressed: mfeldman@stanford.edu \\ $^\text{a}$Department of Biology, Stanford University, Stanford, CA 94305 \\ $^\text{b}$Santa Fe Institute, Santa
Fe, NM 87501    }

\begin{abstract}
\vspace{0.5cm}
\centerline{\textbf{Abstract}}

\noindent Models of conformity and anti-conformity have typically focused on cultural traits with nominal (unordered) variants, such as baby names, strategies (cooperate/defect), or the presence/absence of an innovation. There have been fewer studies of conformity to ``ordinal'' cultural traits with ordered variants, such as level of cooperation (low to high) or fraction of time spent on a task (0 to 1). In these latter studies, conformity is conceptualized as a preference for the mean trait value in a population even if no members of the population have variants near this mean; e.g., 50\% of the population has variant 0 and 50\% has variant 1, producing a mean of 0.5. Here, we introduce models of conformity to ordinal traits, which can be either discrete or continuous and linear (with minimum and maximum values) or circular (without boundaries). In these models, conformists prefer to adopt more popular cultural variants, even if these variants are far from the population mean. To measure a variant's ``popularity'' in cases where no two individuals share precisely the same variant on a continuum, we introduce a metric called $k$-dispersal; this takes into account a variant's distance to its $k$ closest neighbors, with more ``popular'' variants having lower distances to their neighbors. We demonstrate through simulations that conformity to ordinal traits need not produce a homogeneous population, as has previously been claimed. Under some combinations of parameter values, conformity sustains substantial trait variation over many generations. Anti-conformist transmission may produce high levels of polarization.

\vspace{0.3cm}
\centerline{\textbf{Significance Statement}}

\noindent Conformist and anti-conformist biases in acquiring cultural variants have been documented in humans and several non-human species. We introduce a framework for quantifying these biases when cultural traits are \textit{ordinal}, with ordered variants, and either continuous (e.g., level of a behavior) or discrete (e.g., number of displays of a behavior). Unlike previous models, we do not measure a cultural variant’s popularity by its distance to the population mean, but rather by its distance to other cultural variants. We find that conformity can sustain significant trait variation over many generations, challenging the prevailing view that conformity reduces cultural diversity. Anti-conformity may lead to highly polarized or uniformly distributed populations, depending on its strength and on individuals’ precision when copying others.

\vspace{0.3cm}
\noindent Keywords: cultural evolution $|$ conformity $|$ ordinal trait $|$ continuous $|$ discrete 

\end{abstract}

\maketitle

\noindent Cultural traits include information, beliefs, behaviors, customs, and practices that are acquired and transmitted through social mechanisms \cite{CavalliSforzaFeldman:1981, richboyd05}. Such traits may be \textit{nominal} or \textit{ordinal} (Table 1). The variants of nominal cultural traits fall into distinct categories and do not have a natural order or meaningful numerical value, such as baby names, tool types, and the adoption (or not) of a behavior. The variants of ordinal traits are ordered, either in discrete segments or in a continuum and either linearly  or circularly. In the linear case, there are clearly defined smaller and greater values, such as short or long skirts, whereas in the circular case there are not, because the `ends' of the spectrum join together. For example, the time at which an individual chooses to go to bed is circular, selected from a clock in which 11:59 pm on one day and 12:00 am on the next day are very similar to each other.

\bgroup
\def\arraystretch{1.7}%
\begin{table*}
\centering
\caption{Examples of nominal (discrete) and ordinal (discrete or continuous) traits. }
\begin{tabular}{*{4}{l}}
  \toprule
   & Nominal & Ordinal - Linear & Ordinal - Circular \\
  \midrule
  \multirow{3}{*}{\rotatebox[origin=c]{90}{Discrete}}   & \tabitem Cooperate or defect & \tabitem Number of individuals helped &  \tabitem \makecell[l]{Month of the year for an \\ event (e.g., vacation)}  \\
  & \tabitem Vote for party $A$, $B$, or $C$ & \tabitem Education level (high, medium, low)  & \tabitem \makecell[l]{Day of the week for an \\ event (e.g., laundry)}  \\
  & \tabitem Folklore motifs  & \tabitem Number of flowers a bee visits per day  &   \\
    \hline  
    \multirow{3}{*}{\rotatebox[origin=c]{90}{Continuous}} & \hspace{0.1cm} Not applicable. & \tabitem Proportion of resources shared  & \tabitem Color on a color wheel \\
     & & \tabitem Amount of time spent on a task  & \tabitem Choice of bedtime \\ 
    & & \tabitem Skirt length &  \\ 
  \bottomrule   
\end{tabular}
\end{table*}
\egroup

Most previous studies of conformity and anti-conformity have focused on nominal cultural traits that have exactly two variants, known as dichotomous traits. For example, in a study on conformity in nine-spined sticklebacks, the two variants were `swim toward a yellow feeder' and `swim toward a blue feeder'  \cite{PikeLaland:2010}. In an experiment on conformity for mate choice, fruit flies were painted pink or green, and observer fruit flies chose which color to mate with depending on the choices of demonstrator flies \cite{DanchinEtal:2018}. In other experiments on conformity, the variants were the direction, to the left or right, that birds opened a puzzle box slider \cite{AplinEtal:2015a}, and the choice of `same shape' or `different shapes' for people observing two shapes rotated at different angles \cite{MorganEtal:2012}.

For dichotomous cultural traits, the definitions of conformity and anti-conformity are straightforward;  conformity entails that the more popular of the two variants is adopted at a rate \textit{greater} than its frequency, whereas anti-conformity entails that it is adopted at a rate less than its frequency \cite{BoydRicherson:1985}. Boyd and Richerson \cite{BoydRicherson:1985} formalized this definition in a mathematical model, which has been widely used in subsequent theoretical research \cite{HenrichBoyd:1998, HenrichBoyd:2001, Henrich:2001, KamedaNakanishi:2002, Nakahashi:2007, WakanoAoki:2007, MollemanEtal:2013, DentonEtal:2020, DentonEtal:2021, DentonEtal:2021b, BorofskyFeldman:2022}. The model gives rise to the following population dynamics: under conformity, the more popular of the two variants increases in frequency until everyone has it \cite{BoydRicherson:1985}. Under sufficiently low levels of anti-conformity, as in \cite{BoydRicherson:1985}, the two cultural variants stabilize at a population frequency of 50\% each. With high levels of anti-conformity, however, stable cycles between population states, and chaos, may occur \cite{DentonEtal:2020}. 

For a nominal cultural trait with more than two variants (e.g., second and third bullet points in the top-left of Table 1), the definitions of conformity and anti-conformity differ from the two-variant case.  Muthukrishna et al. \cite{MuthukrishnaEtal:2016} pointed out that in a  model with two variants, $A_1$ and $A_2$, the variant $A_1$ must be present at a frequency above 50\% to be the `more popular' variant that is preferred by conformists. However, in a population with four variants, $A_1, A_2, A_3,$ and $A_4$, the variant $A_1$ need only be present at a frequency above $25\%$ to suggest that it is more popular than some of the other variants. They concluded that ``all current models and experiments may have been underestimating the strength of the conformist bias, because there are often more than 2 [variants] in the real world'' (\cite{MuthukrishnaEtal:2016}, p. 9). 

To illustrate the complexity inherent in conformity and anti-conformity when there are more than two variants, consider a sample of $n= 10$ individuals, referred to as ``role models,'' and imagine there are 4 $A_1$, 3 $A_2$, 2 $A_3$, and 1 $A_4$. Would a conformist adopt, say, $A_2$ at a rate greater than its frequency or less than its frequency? In other words, is $A_2$ `popular' or `unpopular'? To answer this question, Denton et al. \cite{DentonEtal:2022} denoted by $r$ the number of distinct cultural variants in the sample of $n$, and classified a variant as `popular' if it occurs in more than $\tfrac{n}{r}$ role models. In this case, $\tfrac{n}{r}$ is $\tfrac{10}{4} = 2.5$, so variants $A_1$ and $A_2$ would be considered popular while $A_3$ and $A_4$ are unpopular. Further, in \cite{DentonEtal:2022}, a conformist is more likely to acquire $A_1$ than $A_2$ because $A_1$ is more popular than $A_2$, while an anti-conformist is more likely to acquire $A_4$ than $A_3$ because $A_4$ is rarer than $A_3$. A general formula for conformity and anti-conformity to $n$ role models in a population with $m$ variants is given by Eq. (17) in \cite{DentonEtal:2022}. 

The population dynamics that occur in the model of \cite{DentonEtal:2022} for a nominal trait with $m \geq 2$ cultural variants are as follows (see their Table 2 for a summary and a comparison to the two-variant case). Under conformity, the $\ell \geq 1$ cultural variant(s) that are initially most popular (i.e., have a higher frequency than all others) increase in frequency until they reach an equilibrium of $\tfrac{1}{\ell}$ and all other variants disappear. Under sufficiently low levels of anti-conformity, similar to the two-variant model \cite{BoydRicherson:1985}, the frequencies of all $m$ variants present in the population reach frequencies of $\tfrac{1}{m}$, whereas with sufficiently high levels of anti-conformity, cycles and chaos are possible.

In comparison with research on nominal cultural traits, fewer studies have considered ordinal cultural traits. An ordinal trait that was linear and continuous was modelled in \cite{CavalliSforzaFeldman:1973}, \cite{SmaldinoEpstein:2015}, and \cite{MorganThompson:2020}, where group-mean transmission, referred to as ``conformity'' by \cite{SmaldinoEpstein:2015} and \cite{MorganThompson:2020}, entailed adopting a cultural variant equal to or near the group's average value. In  \cite{SmaldinoEpstein:2015}, the population could converge to an equilibrium in which everyone shared the same variant, even if individuals were not entirely conformist but adopted variants that deviated from the population mean (i.e., had a ``distinctiveness preference''). 

However, defining conformity in relation to the mean variant in a population departs significantly from the definition of conformity in nominal models. Consider an example in which 5 role models are sampled, and 3 individuals are wearing long dresses while 2 are wearing short dresses. In the nominal, two-variant model of conformity, conformists would be more likely to adopt the long dress than the short dress, and could never adopt a medium-length dress (not observed in any individual). However, in  \cite{SmaldinoEpstein:2015} and \cite{MorganThompson:2020}, conformists prefer the average, medium-length dress over either a long or a short dress. Here, we develop models of conformity to ordinal traits that do not prioritize the group mean but preserve the idea that a conformist adopts a popular variant. We consider both discrete and continuous traits, as well as linear and circular traits.

In a continuous-trait model, it is highly unlikely that two or more role models will have precisely the same variant. Therefore, `popular' cultural variants cannot be defined as being present in many role models. Instead, we define popular cultural variants with reference to their \textit{distance} from one another, and introduce a measure that we call $k$-dispersal to quantify the popularity of a variant in a continuous-trait model. Below, we describe this metric as well as other features of our model, first by introducing models of linear ordinal traits and subsequently extending these to incorporate circular ordinal traits. Finally, the population dynamics of these models are explored using simulations. \vspace{-0.4cm}

\section{Methods} \vspace{-0.1cm}

\subsection{Model assumptions and overview} \vspace{-0.2cm}

\noindent Our model includes the following commonly made assumptions (e.g., see \cite{BoydRicherson:1985, DentonEtal:2022}). First, there are discrete and non-overlapping generations of parents and offspring. Second, each offspring takes a random sample of $n$ ``role models'' from the parent population, known as oblique cultural transmission. They could also sample from other offspring within their generation, known as horizontal cultural transmission \cite{CavalliSforzaFeldman:1981}, but as in previous models \cite{BoydRicherson:1985, DentonEtal:2022}, oblique and horizontal transmission are mathematically equivalent here. For horizontal transmission to occur, members of the offspring generation must \textit{already have} cultural variants that can be transmitted, and a natural assumption is that each offspring initially inherits the cultural variant of one of its parents, that is, by vertical transmission. Therefore, the frequencies of cultural variants in the parental generation and the initial, vertically transmitted frequencies of cultural variants in the offspring are the same, so the consequences of oblique and horizontal transmission are identical. 

Unlike many previous models of conformity, which assume an infinite population size \cite{BoydRicherson:1985, DentonEtal:2020, DentonEtal:2021, DentonEtal:2022} (but see \cite{LappoEtal:2023}), we assume a finite population size. Before detailing the rules that individuals use to acquire a cultural trait, the following is a brief overview of the entire model:
\begin{enumerate}
    \item A population comprises $N$ individuals, each of which  possesses a cultural variant represented by a value ranging between 0 and 1 (without loss of generality). 
    \item Members of this population `reproduce' to generate $N$ offspring, which initially do not have cultural variants. 
    \item Then, \textit{each} of the $N$ offspring acquires a cultural variant through the following steps:
    \begin{enumerate}
        \item An offspring randomly samples $n$ role models from the previous generation. 
        \item From these $n$ role models, an offspring generates a probability distribution, which is based on its level of conformity, anti-conformity, or a third, unbiased type of frequency-dependent transmission known as random copying  (these are described in detail in Sections B and C).
        \item From this customized probability distribution, an offspring adopts a cultural variant. 
    \end{enumerate}
\end{enumerate}
Steps 1 through 3 are repeated, with the  offspring population becoming the new adult population in each generation. A diagram of the model overview is in Supplementary Information 1, Figure S1.1. \vspace{-0.4cm}

\subsection{Models of linear ordinal traits with \\ continuous variation}  \vspace{-0.2cm}

\noindent Here, we  introduce models of frequency-dependent transmission (namely conformity, anti-conformity, and random copying) of linear traits, providing details about Step 3 from Section A above.  Before introducing models of conformity and anti-conformity, we discuss the baseline model of unbiased frequency-dependent transmission, namely random copying.  \vspace{-0.4cm}
\subsubsection{Random copying}  \vspace{-0.3cm}

\noindent In previous studies on nominal cultural traits, random copying occurs when individuals ``choose another member of the previous generation at random and copy their cultural trait'' \cite{MesoudiLycett:2009} (p. 42). For continuous traits such as dress length, however, it is unlikely that one individual would copy the exact length in millimeters of another individual's dress. Instead, we define random copying as follows. Assume, without loss of generality, that the continuous cultural trait has a minimum of 0 and a maximum of 1, and for a linear trait the minimum and maximum are not joined together. The circular trait analog of this model, where the minimum and maximum are joined, will be discussed in Section D. 

\begin{definition}\emph{Random copying for a continuous linear trait.}
    An individual (a) samples the cultural variant of a member of the previous generation at random, (b) sets this value as the mean of a normal distribution with standard deviation $\sigma$, (c) re-normalizes this probability distribution so that the minimum trait value is 0 and the maximum trait value is 1, and (d) samples a value at random from this distribution. 
\end{definition}

\begin{figure}[ht]
\centering
\includegraphics[width=0.5\textwidth]{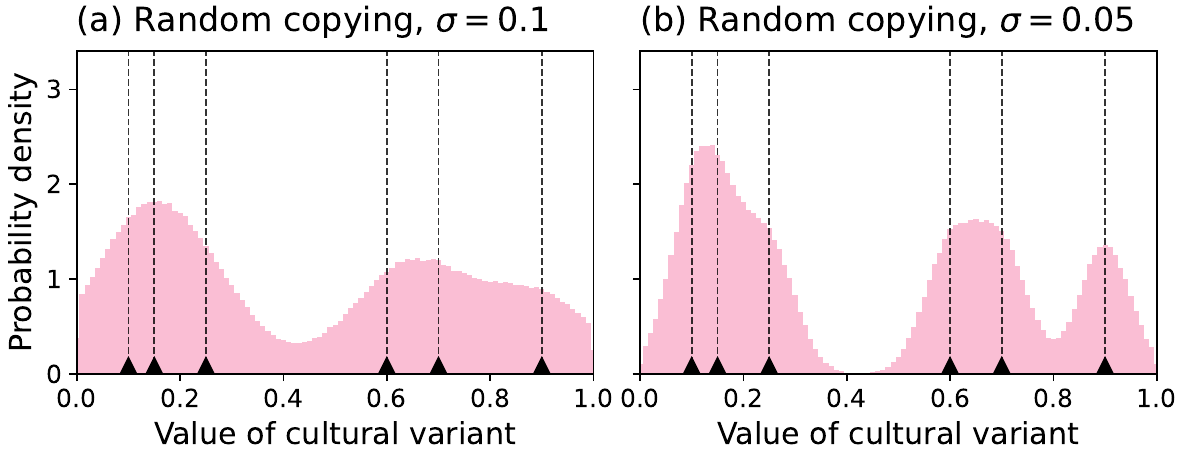} 
    \caption{A single individual's probability density of adopting a variant given a role model sample, assuming random copying according to Definition 1. A sample of $n=6$ role models with variants $\boldsymbol{x} = (0.1,0.15,0.25,0.6,0.7,0.9)$ is shown in black triangles and dashed lines. The curve shows the approximate probability density of adopting each cultural variant given this sample $\boldsymbol{x}$, obtained through performing 1,000,000 replications of Definition 1. (In whole-population simulations of our model, which will be shown in the Results section, each individual will actually perform only one iteration of Definition 1 to acquire a single cultural variant; thus, these 1,000,000 iterations are for illustrative purposes only.) (a) $\sigma = 0.1$, (b) $\sigma = 0.05$ in Definition 1. Histogram bin widths are 0.01. } \label{variant_adopt}
\end{figure}

Note that Definition 1 is effectively a kernel density estimation (e.g., see Figure 1 in \cite{Weglarczyk:2018}). Two examples of the random copying model  in Definition 1 with different values of $\sigma$ are given in Figure \ref{variant_adopt}. This figure shows the probability density of acquiring a cultural variant for a \textit{single} individual in a single generation, given a particular sample $\boldsymbol{x}$ of role models' cultural variants. Specifically, there are $n=6$ role models with variants $\boldsymbol{x} = (0.1,0.15,0.25,0.6,0.7,0.9) $ shown on the $x$-axis of Figure \ref{variant_adopt} as  black triangles. Repeating the steps in Definition 1 a large number of times (i.e., 1,000,000), the figure illustrates the probability density at which the focal individual will adopt the range of possible cultural variants. This figure does not depict the evolutionary process, but rather the mechanism by which an individual adopts a trait value in a single instance. In the full evolutionary model (see Results), every individual in the population of size $N$ will independently draw their own sample of $n$ role models from the previous generation and construct their own customized probability distribution according to Definition 1.

\vspace{-0.8cm}

\subsubsection{Conformity and anti-conformity}  \vspace{-0.3cm}

\noindent Past studies on nominal traits have defined conformity as the tendency to adopt a more common variant with a probability greater than its frequency, and a less common variant with a probability less than its frequency \cite{BoydRicherson:1985, DentonEtal:2020, DentonEtal:2022}. For continuously varying ordinal cultural traits, however, it is virtually impossible that two or more individuals have exactly the same variant. Therefore, ``commonness,'' namely the count of individuals with a particular variant, is a poor indicator of a variant's popularity. Instead, we consider the \textit{proximity} of cultural variants to other variants. For instance, if $n=5$ people's hair lengths were observed and took the values (in inches)
$ \boldsymbol{x} = (5, 12, 12.5, 13, 20),$
then, although each variant has a frequency of $\tfrac{1}{n} = \tfrac{1}{5}$, a conformist would be more likely to adopt a variant around 12-13 inches than around 5 or 20 inches, as the latter are more dispersed from the other variants.

Note, however, that an individual adopting the ``random copying'' strategy also has a higher probability of adopting variants in the proximity of less dispersed sampled variants than more dispersed ones (e.g., Figure \ref{variant_adopt}). The difference between conformity and random copying is that a conformist is \textit{disproportionately} more likely to adopt the less dispersed than the more dispersed variants relative to the strategy of randomly choosing a value to be the mean of a normal distribution (as in Definition 1). First, we quantify dispersal:

\begin{definition} \emph{$k$-dispersal for a linear trait.} \label{k_dispersal}
    In a sample of $n$ role models $\boldsymbol{x} = (x_1,x_2,\dots,x_n)$, consider variant $x_i$. The $k$-dispersal of variant $x_i$ is the sum of the $k$ shortest absolute distances between variant $x_i$ and the other variants in $\boldsymbol{x}$. Note that $k<n$.
\end{definition}

For example, let $ \boldsymbol{x} = (5, 12, 12.5, 13, 20)$. The variant 5 has a $k$-dispersal of $12-5=7$ if $k=1$; 14.5 if $k=2$; and so on. In contrast, the variant $12.5$ has a $k$-dispersal of 0.5 if $k=1$ and 1 if $k=2$. Thus, the values around 12-13 have lower $k$-dispersals than the values 5 or 20. 

Without loss of generality, assume again that the continuous linear trait lies within the values 0 and 1. The maximum $k$-dispersal for a variant $x_i$ from the sample of $n$ role models occurs if variant $x_i$ is at one extreme (0 or 1) and all of the other $n-1$ variants in the sample are at the other extreme (1 or 0). The minimum $k$-dispersal for a variant $x_i$ occurs if all variants are identical. (Even if some or all variants are identical, they are still represented as $\boldsymbol{x} = (x_1,x_2,\dots,x_n)$ because here the subscript $i$ in $x_i$ refers to the individual observed, not to the type of variant.) Denote the $k$-dispersal of variant $x_i$ in a sample of $\boldsymbol{x} = (x_1,x_2,\dots,x_n)$ by $f_{i,k}( \boldsymbol{x})$. Then $\max f_{i,k}(\boldsymbol{x}) = k$, $\min f_{i,k}(\boldsymbol{x}) = 0$, and $\text{mean } \overline{f_{k}}(\boldsymbol{x}) = \sum_{i=1}^n f_{i,k}(\boldsymbol{x}) / n$. 

For a given value of $k$, let the probability of selecting variant $x_i$ (which will become the mean of a normal distribution) from a sample $\boldsymbol{x}$ of role models be 
\begin{align} \label{eq.prob.xi}
    P_k(x_i \mid \boldsymbol{x}) &= \frac{1}{n} + \frac{g_{i,k}(\boldsymbol{x})d_k(\boldsymbol{x})}{n}
\end{align}
(similar to Eq. (17a) in ref. \cite{DentonEtal:2022}). With random copying, the conformity coefficient $d_k(\boldsymbol{x})$ is zero, in which case \mbox{$P_k(x_i \mid \boldsymbol{x}) = \frac{1}{n}$}. This means that each of the $n$ randomly sampled role models is selected with probability $\tfrac{1}{n}$, equivalent to randomly sampling one individual from the previous generation (Definition 1). With conformity, $d_k(\boldsymbol{x}) > 0$, and with anti-conformity, $d_k(\boldsymbol{x}) < 0$.

Now, we must define $g_{i,k}(\boldsymbol{x})$ in such a way that if variant $x_i$ is ``less dispersed,'' this value is positive, whereas if $x_i$ is ``more dispersed,'' this value is negative (similar to the definition for nominal traits in \cite{DentonEtal:2022} but substituting ``less dispersed'' for ``more frequent''). Note that $g_{i,k}(\boldsymbol{x})$ does not appear in models of random copying, but only in conformity and anti-conformity models.  Let  
\small 
\begin{align} \label{eq.g}   
g_{i,k}(\boldsymbol{x}) &= \begin{cases} 
    -\frac{f_{i,k}(\boldsymbol{x})}{\sum\limits_{z \in \ROMAN{I}}z} \hspace{0.9cm} \text{if } f_{i,k}(\boldsymbol{x}) \in \ROMAN{I}, \ \ROMAN{I} = \{y: y> \overline{f_k}(\boldsymbol{x})  \}\\ 
    \frac{ [f_{i,k}(\boldsymbol{x})]^{-1} }{\sum\limits_{z \in \ROMAN{II}} z^{-1}} \hspace{0.7cm} 
    \begin{aligned} &\text{if } f_{j,k}(\boldsymbol{x}) > 0 \ \forall j \text{ and }  f_{i,k}(\boldsymbol{x}) \in \ROMAN{II}, \\
    &\ROMAN{II} = \{y: 0 < y < \overline{f_k}(\boldsymbol{x}) \} 
    \end{aligned} \\
    \frac{1}{\sum_{j} [f_{j,k}(\boldsymbol{x}) = 0]}  \hspace{0.2cm} \text{if } f_{i,k}(\boldsymbol{x}) = 0 < \overline{f_k}(\boldsymbol{x}) \\ 
    0 \hspace{1.9cm} \text{if } \exists j,   f_{j,k} (\boldsymbol{x}) = 0 < f_{i,k}(\boldsymbol{x}) < \overline{f_k}(\boldsymbol{x}) \\
    0 \hspace{1.9cm}   \text{if } f_{i,k} (\boldsymbol{x}) = \overline{f_k}(\boldsymbol{x}). 
\end{cases}
\end{align} \normalsize 
If the dispersal of a variant $x_i$ is higher than the average dispersal of variants in the sample, i.e., $f_{i,k}(\boldsymbol{x}) > \overline{f_k}(\boldsymbol{x})$, then the value of $g_{i,k}(\boldsymbol{x})$ will be negative, and this value will become more negative as the dispersal $f_{i,k}(\boldsymbol{x})$ increases (row 1 in Eq. \ref{eq.g}). Therefore, an anti-conformist with $d_k(\boldsymbol{x}) < 0$ will have a higher probability of adopting a cultural variant as its dispersal increases (Eq. \ref{eq.prob.xi}).

To understand row 2 in Eq. (\ref{eq.g}),  consider a case in which $f_{j,k}(\boldsymbol{x}) > 0$ for all $j$, and variant $x_i$ is less dispersed than the average dispersal in the sample, so $0<f_{i,k}(\boldsymbol{x}) < \overline{f_{k}}(\boldsymbol{x})$. In this case, $g_{i,k}(\boldsymbol{x})$ will be positive, and it will increase as the dispersal $f_{i,k}(\boldsymbol{x})$ decreases. Therefore, a conformist with $d_k(\boldsymbol{x}) > 0$ will have a higher probability of adopting a cultural variant as its dispersal from other variants decreases (Eq. \ref{eq.prob.xi}).

If $x_i$ is extremely close to other cultural variants such that its dispersal $f_{i,k}(\boldsymbol{x}) \xrightarrow[]{} 0$, then the numerator in row 2 of Eq. (\ref{eq.g}) will be very large, as will at least one value within the sum in the denominator (as this sum contains the value of the numerator). If $f_{i,k}(\boldsymbol{x}) = 0$, meaning that some cultural variants in the sample $\boldsymbol{x}$ are identical, then the expression in row 2 cannot be applied because it would entail division by 0; instead, row 3 in Eq. (\ref{eq.g}) is applied. The square brackets in row 3 are Iverson brackets, meaning that the expression is 1 if the statement within the brackets is true and 0 otherwise. Therefore, row 3 of Eq. (\ref{eq.g}) captures the idea that if $m$ cultural variants have $k$-dispersals of 0, they each will have the same $g_{i,k}(\boldsymbol{x})$ value, namely $\tfrac{1}{m}$. Rows  4 and 5 of Eq. (\ref{eq.g}) ensure that $\sum_{i=1}^n g_{i,k}(\boldsymbol{x}) = 0$ so that in Eq. (\ref{eq.prob.xi}), $\sum_{i=1}^n P_k(x_i \mid \boldsymbol{x}) = 1$.

Conformity and anti-conformity can now be defined in terms of $n$, the number of role models; $k$, the value used in Definition 2; $\sigma$, the standard deviation; and $d_k(\boldsymbol{x})$, the function in Eq. (\ref{eq.prob.xi}) that gives the strength of (anti-) conformity.

\begin{definition}\emph{Conformity and anti-conformity to a continuous linear trait.} \label{conformity}
    An individual samples at random $n$ role models from the previous generation, and observes their cultural variants $\boldsymbol{x} = (x_1,x_2,\dots,x_n)$. For a specified $k$, the individual selects value $x_i$ in $\boldsymbol{x}$ with a probability given by Eq. (\ref{eq.prob.xi}). This value is set as the mean of a normal distribution with standard deviation $\sigma$, which is then re-normalized to have a minimum of 0 and a maximum of 1; finally, the individual samples a value at random from this distribution. There is conformity if $d_k(\boldsymbol{x}) > 0$ and anti-conformity if  $d_k(\boldsymbol{x}) < 0$ in Eq. (\ref{eq.prob.xi}). 
\end{definition}

\begin{figure}[ht]
\centering
\includegraphics[width=0.5\textwidth]{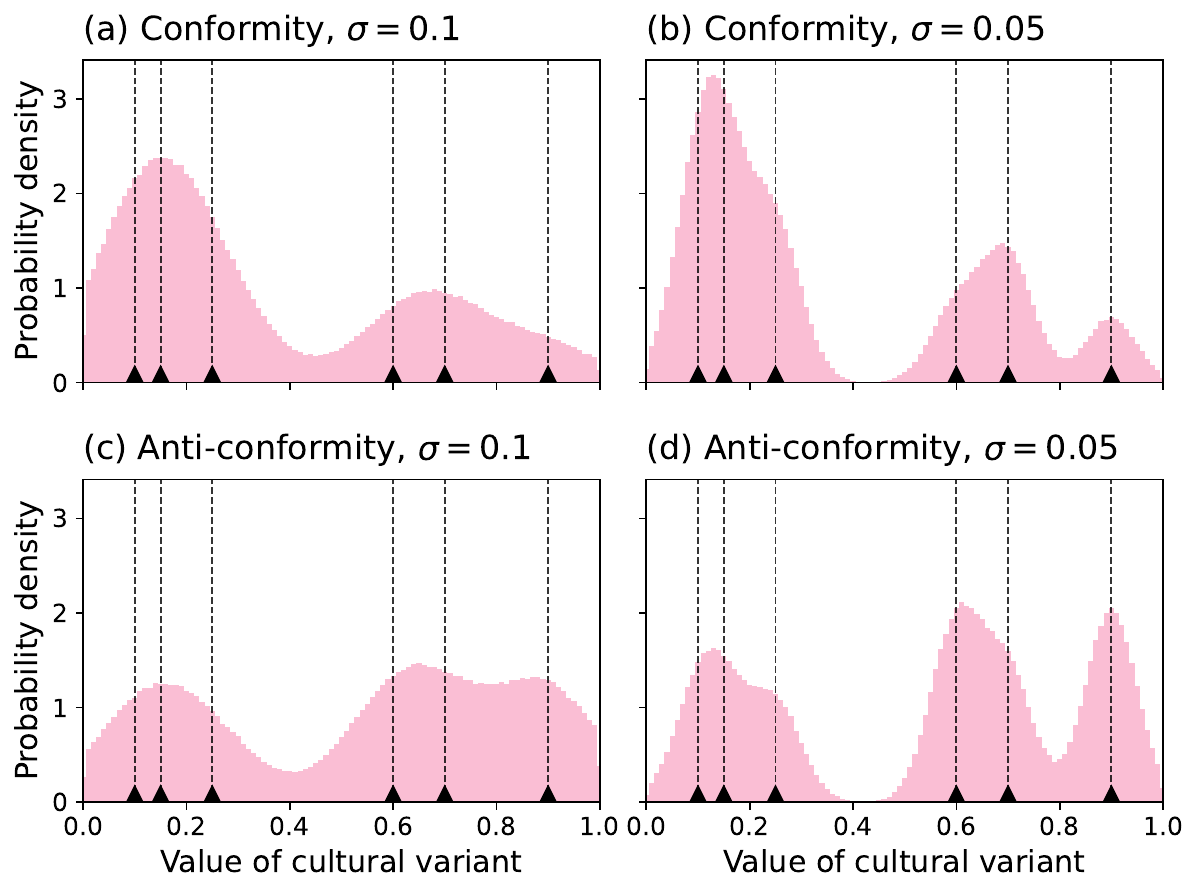} 
    \caption{An individual's probability density $P_k(x_i \mid \boldsymbol{x})$ of adopting a linear, continuous cultural variant given a role model sample, with $k=2$ and either conformity ($d_k(\boldsymbol{x}) = 0.9$, top row) or anti-conformity ($d_k(\boldsymbol{x}) = -0.9$, bottom row). The sample of $n=6$ role models with variants $\boldsymbol{x} = (0.1,0.15,0.25,0.6,0.7,0.9)$ is shown in black triangles and dashed lines. The curve shows the approximate probability density for a single individual to adopt a variant given this sample, calculated by taking 1,000,000 independent samples following Definition 3. (a,c) $\sigma = 0.1$, (b,d) $\sigma = 0.05$.  Histogram bin widths are 0.01.} \label{adopt_conformity_1}
\end{figure}

In SI 2, the upper and lower bounds of the conformity coefficient $d_k(\boldsymbol{x})$ are calculated.  Figure \ref{adopt_conformity_1} shows the probabilities of choosing a cultural variant given the role model sample $\boldsymbol{x} = (0.1,0.15,0.25,0.6,0.7,0.9)$, with either conformity (top row) or anti-conformity (bottom row). Compared to the case of random copying shown in Figure \ref{variant_adopt}, conformity makes the tall peaks taller and the short peaks shorter, whereas anti-conformity has the opposite effect.

\begin{figure}[ht]
\centering
\includegraphics[width=0.5\textwidth]{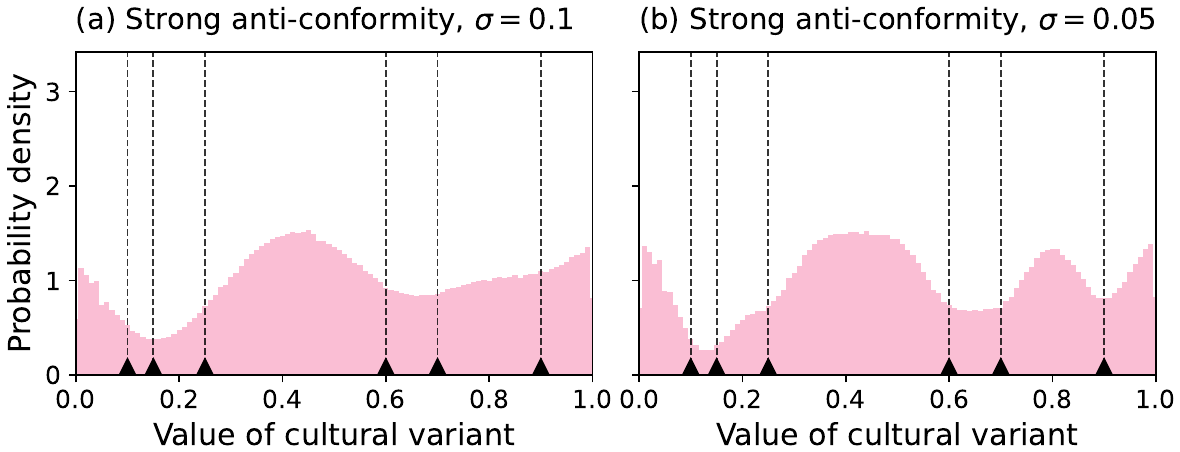} 
    \caption{Probability density of adopting cultural variants given a role model sample under strong anti-conformity. Under strong anti-conformity, an individual is qualitatively ``repelled'' by its role models. In this model, a probability density function following Definition 1 is obtained, ``flipped'' vertically, and normalized. In the figure, an example set of $n=6$ role models with variants $\boldsymbol{x} = (0.1,0.15,0.25,0.6,0.7,0.9)$ is marked with black triangles and vertical dashed lines. The curve shows the probability density function for a single individual to adopt a variant given the example role models. (The figure is computed by taking 1,000,000 independent samples from the probability distribution described in SI 3). (a) $\sigma = 0.1$, (b) $\sigma = 0.05$.} \label{super_anticonform}
\end{figure}

Figure \ref{adopt_conformity_1} captures the idea that anti-conformists have a higher probability of copying cultural variants seen in some role models (namely, individuals with ``unpopular'' cultural variants) relative to the random copying strategy (Figure \ref{variant_adopt}), as is the case for many studies of nominal traits \cite{BoydRicherson:1985, HenrichBoyd:1998, Henrich:2001, HenrichBoyd:2001, KamedaNakanishi:2002, Nakahashi:2007, WakanoAoki:2007, MollemanEtal:2013, BorofskyFeldman:2022, DentonEtal:2020, DentonEtal:2021, LappoEtal:2023, DentonEtal:2021b, DentonEtal:2023}. However, this is not the only possible conceptualization of anti-conformity. In Figure \ref{adopt_conformity_1}c,d, the probability of adopting a variant not seen in any role models (e.g., 0.4) is fairly low. Efferson et al. \cite{EffersonEtal:2008} introduced a model of ``strong anti-conformity'' to a dichotomous nominal trait in which the probability of adopting a cultural variant that is not present in any role models is very high, and we generalize this idea to an ordinal trait in Figure \ref{super_anticonform}. Rather than adjusting the peaks of Figure \ref{variant_adopt}, we \textit{invert} the distribution of Figure \ref{variant_adopt} (details in SI 3).  Because this computation involves sampling from a customized probability density function $y=\rho(v)$ in which each value, $y$, is inverted across a horizontal axis, we use a discretized rather than a continuous probability distribution (with a small separation between discrete values to approximate a continuous model, namely cultural variants $0, 0.001, 0.002, \dots, 1$; see SI 3). \vspace{-0.4cm}

\subsection{Models of linear ordinal traits with \\ discrete variation} \vspace{-0.2cm}

\noindent Consider an ordinal trait that can take discrete values $\alpha_1, \alpha_2, \dots, \alpha_{\ell}$. For example, a bee might visit $0, 1, 2, \dots, 100$ flowers in a given day (but it could not visit, say, $2.75$ flowers). Hence, $\alpha_1 = 0, \alpha_2 = 1, \dots \alpha_{101} = 100$. 

For the discrete versions of each of our models presented in Section B (Figures \ref{variant_adopt}-\ref{super_anticonform}), an individual first selects \textit{any} cultural variant, following the previously mentioned protocols. For example, a conformist would sample $n$ role models, observe their cultural variants $\boldsymbol{x}$, and, for a specified $k$, select a value $x_i$ following Eq. (\ref{eq.prob.xi}). Then, the individual rounds $x_i$ to the nearest value $\alpha_1, \alpha_2, \dots, \alpha_{\ell}$. Analogs of Figures \ref{variant_adopt}-\ref{super_anticonform} with 21 cultural variants are given in SI 4 (Figures S4.1-S4.3, respectively).  \vspace{-0.4cm}

\subsection{Models of circular ordinal traits}  \vspace{-0.2cm}

\begin{figure}[ht]
\centering
\includegraphics[width=0.5\textwidth]{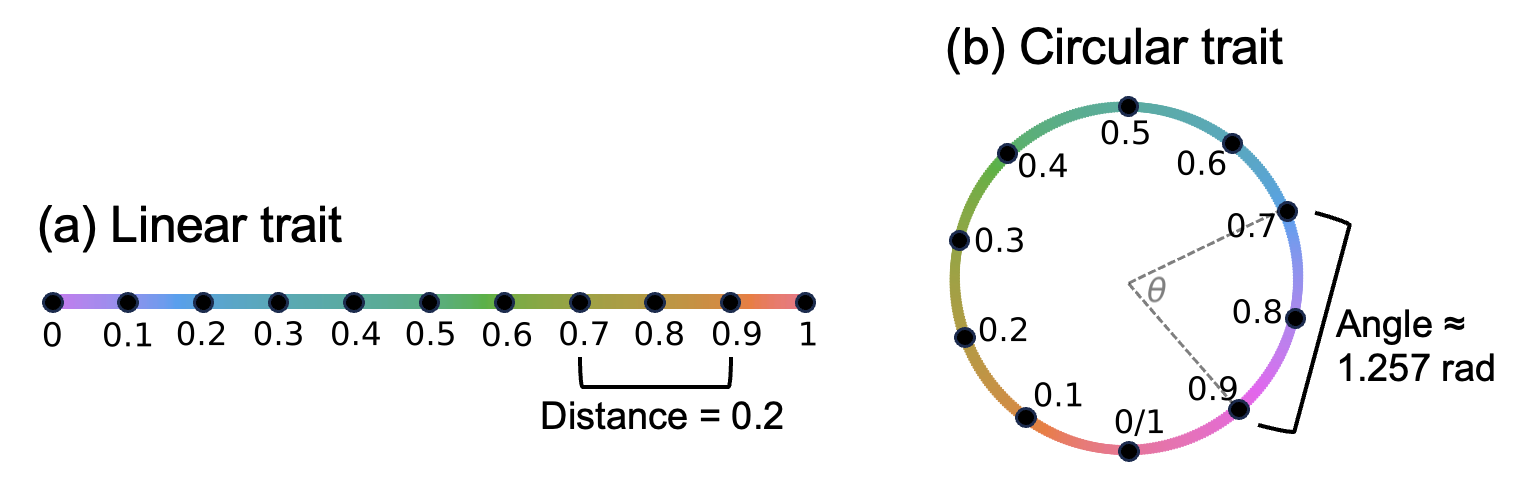} 
    \caption{Example calculations of dispersal between two variants, 0.7 and 0.9, for a linear (left) or circular (right) trait. In practice, dispersals would be calculated between all variants $x_1,x_2,\dots,x_n$ in a sample of $n$ role models prior to applying Definition 2.}  \label{linear.vs.circular}
\end{figure}

\noindent Models of circular traits differ from  the above models of continuous and discrete linear traits in the following ways:
\begin{enumerate}
    \item In modifying Definitions 1 and 3, a role model's variant, $x_i$, is set as the mean of a normal distribution with standard deviation $\sigma$ that is \textit{not} restricted to $[0,1]$. If a value above 1 is sampled, it is ``wrapped around'' to the other end of the distribution, e.g., $1.1$ becomes $0.1$ (Figure \ref{linear.vs.circular}b). Similarly, values below 0 are wrapped around with $-0.1$ becoming $0.9$, for example. 
    \item In modifying Definition 2, ``$k$ shortest absolute distances'' is replaced by ``$k$ smallest angles, in radians'' when cultural variants are represented in a circular configuration. Specifically, this angle is calculated as follows. The circle of cultural variants that lie within $[0,1]$ has a circumference of $1$ and a radius of $r = \tfrac{1}{2\pi}$. The smallest arc length between two points $x_i$ and $x_j$ on the circle is $s = \min(|x_i-x_j|, 1 - |x_i-x_j|)$. Then, the angle between $x_i$ and $x_j$ is $\theta = \frac{s}{r}$. An example is shown in Figure \ref{linear.vs.circular}b. 
    \item The model of strong anti-conformity for a linear trait (Figure \ref{super_anticonform}) entails vertically ``flipping'' the distribution from Definition 1 so the role models qualitatively repel, rather than attract, new individuals. For strong anti-conformity to a circular trait, the distribution is ``flipped'' and also ``wrapped around'' the circle, as in Figure \ref{linear.vs.circular}b. In the circular case, role models repel a new individual, and the repulsion wraps around in the sense that a role model at $x_i=0.001$ will repel new individuals from choosing $x_j=0.0$, and also from choosing $x_j=1.0$. For quantitative details of strong anti-conformity (for both linear and circular traits), see SI 3.
\end{enumerate}

Figures S5.1-S5.6 in Supplementary Information 5 show circular-trait analogs of Figures \ref{variant_adopt}-\ref{super_anticonform} and S4.1-S4.3. \vspace{-0.2cm}

\section{Results of Evolution} \vspace{-0.1cm}

\noindent The parameters of our model are: $n$, the number of sampled role models per individual; $\sigma$, the standard deviation of the normal distribution in Definitions 1 and 3, or ``precision'' in trait adoption;  $k$ values used in the measure of $k$-dispersal (Definition 2); conformity coefficients $d_k(\boldsymbol{x})$; population size, $N$; and number of generations for which simulations are run. Here, and in SI 6, the effects of different model parameters on the population dynamics are described.

\begin{figure*}[t]
\centering
\includegraphics[width=12.5cm,height=9.4cm]{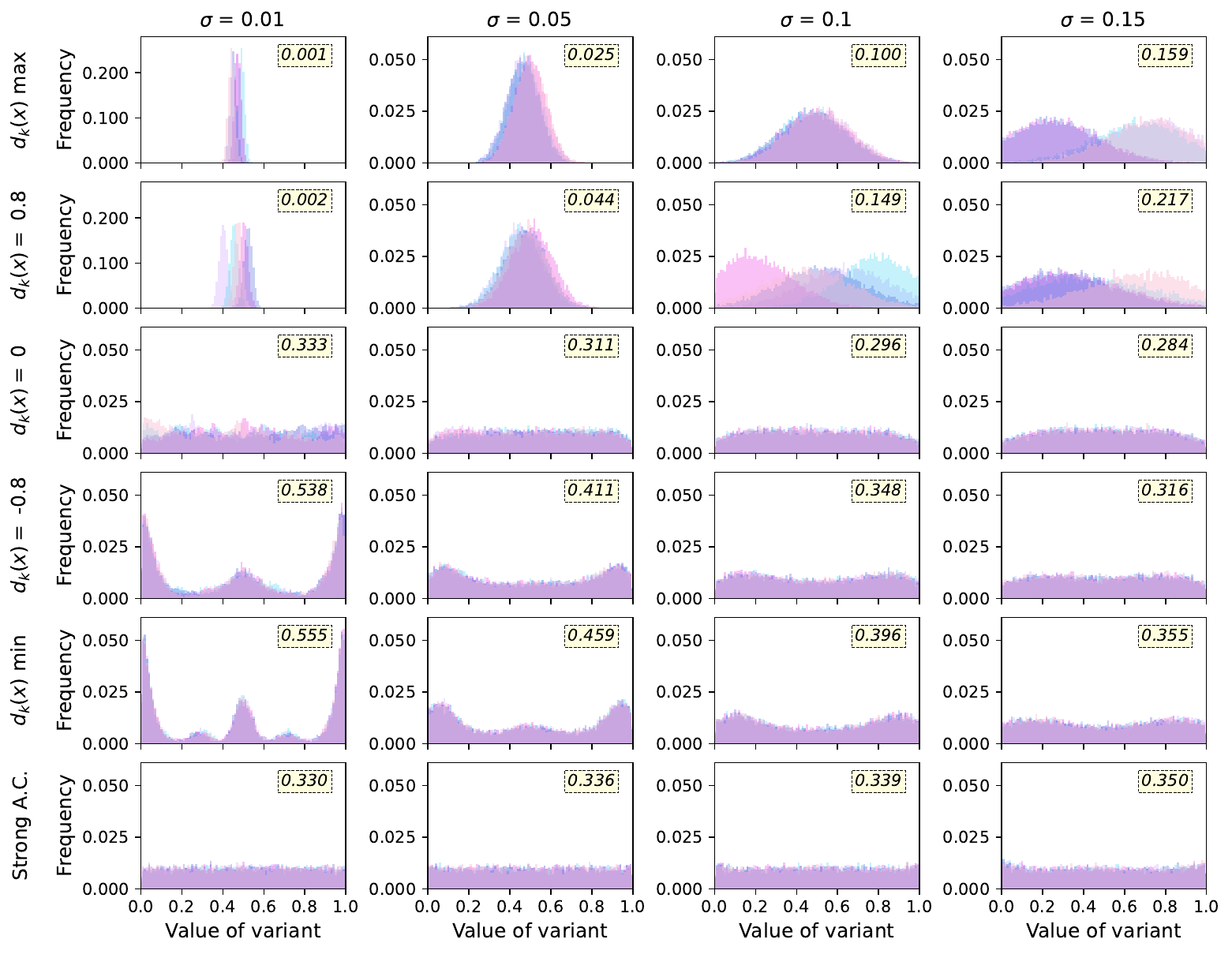}
\caption{Population distributions after 100 generations for a continuous linear trait (rows 1-5 from the top) or nearly continuous linear trait (row 6, with ``Strong A.C.'' standing for strong anti-conformity; see SI 3), with $k=2$, $n=5$, and $N=10,000$. 5 replications were performed, each beginning from the same initial population (top right of Figure S1.1a, which was generated by sampling from a uniform distribution on $[0,1]$). Each of the 5 final distributions is plotted in a different color. For each of these 5 final population distributions with cultural variants $\boldsymbol{x} = (x_1,x_2,\dots,x_N)$, the polarization index  $F = \frac{\text{Var}(\boldsymbol{x})}{\text{Mean}(\boldsymbol{x}) \cdot [1 - \text{Mean}(\boldsymbol{x})]}$ (see \cite{LeimarEtal:2022}) is calculated, and then the 5 $F$ values are averaged to produce the value shown in a yellow box in the top-right corner of each panel (rounded to 3 decimal places).  Values of $d_k(\boldsymbol{x})$ and $\sigma$ are in row and column labels, respectively. The label ``$d_k(\boldsymbol{x}) \min$'' refers to the case in which, for each $\boldsymbol{x}$, $d_k(\boldsymbol{x})$ is set to its lower bound (SI 2) plus $10^{-5}$. Similarly, ``$d_k(\boldsymbol{x}) \max$'' means that  $d_k(\boldsymbol{x})$ is set to its upper bound minus $10^{-5}$ (SI 2).   Histogram bin widths are 0.01, and $y$-axes differ in some of the panels (e.g., left-most column's top two panels).}\label{linear.sigma.d.fixed.n}
\end{figure*}

In Figure \ref{linear.sigma.d.fixed.n}, parameters $k$, $n$, and $N$ are held constant at 2, 5, and 10,000, respectively, and the effects of $d_k(\boldsymbol{x})$ (rows) and  $\sigma$ (columns) on the population  are shown after 100 generations for 5 different replicates, each of which is displayed in a different color. Figure S6.1 in SI 6 shows the same populations after 1000 generations. In Figures \ref{linear.sigma.d.fixed.n} and S6.1, the cultural variants in all initial populations were sampled from a uniform distribution on $[0,1]$ and are shown in Figure S1.1a (i.e., all initial populations are the same).

Moving from left to right across any row in Figure \ref{linear.sigma.d.fixed.n} or Figure S6.1, increasing the standard deviation, $\sigma$, has a smoothing effect that results in peaks and valleys in the population distributions becoming less pronounced. This is intuitive, as increasing $\sigma$ decreases the precision with which a given role model's cultural variant is acquired, introducing more randomness and less clustering of cultural variants in the population. 

Moving from the top row to the fifth row in Figure \ref{linear.sigma.d.fixed.n}, i.e., decreasing the level of conformity $d_k(\boldsymbol{x})$ in Eq. (\ref{eq.prob.xi}), we see that populations become more ``polarized''. Following Leimar et al. \cite{LeimarEtal:2022}, we quantify the level of polarization in a population of size $N$ with cultural variant distribution $\boldsymbol{x} = (x_1, x_2, \dots, x_N)$ at generation 100 as $F = \frac{\text{Var}(\boldsymbol{x})}{\text{Mean}(\boldsymbol{x}) \cdot [1 - \text{Mean}(\boldsymbol{x})]}$ (shown in yellow boxes in the top-right corners of the panels in Figure \ref{linear.sigma.d.fixed.n}). The intuition behind this formula is that if all members of the population share the same cultural variant, then $F = 0$ (no polarization), whereas if each member of the population has one of the extreme values ($x_i = 0$ or $x_i = 1$) and not all members have the same variant, then $F = 1$.  

In each panel of Figure \ref{linear.sigma.d.fixed.n}, five different replicates of the model are shown. The value of $F$ is calculated \textit{separately} for each of these independent runs of the model, and then averaged to get the final value shown in the top-right corner of each panel. Thus, for example, the top-right panel of Figure \ref{linear.sigma.d.fixed.n} has a lower $F$ value than the other panels in the same column because $F$ is calculated separately after each (differently colored) complete iteration (100 generations), even though across iterations in the top-right panel, there is more variance (with some runs moving toward the left and some toward the right). In contrast, in the far left, second from the bottom panel in Figure \ref{linear.sigma.d.fixed.n}, all 5 replicates almost entirely overlap with one another and span a wide range of values, so the average $F$ value is larger.

With random copying (third row from the top in Figure \ref{linear.sigma.d.fixed.n}) or anti-conformity (bottom three rows), the variance in trait distribution is large after 100 generations. This is true even though all individuals in the population have the \textit{same} levels of conformity, e.g., $d_k(\boldsymbol{x})$ with $k=2$ in rows 3-5 (the results are similar with $k=3$; Figure S6.2). In contrast, in the model of Smaldino and Epstein \cite{SmaldinoEpstein:2015} when all individuals shared the same levels of conformity or anti-conformity, the variance of the final distribution of cultural variants was zero.

\begin{figure*}
\raggedright
  \centering
  \includegraphics[width=12.5cm,height=9.4cm]{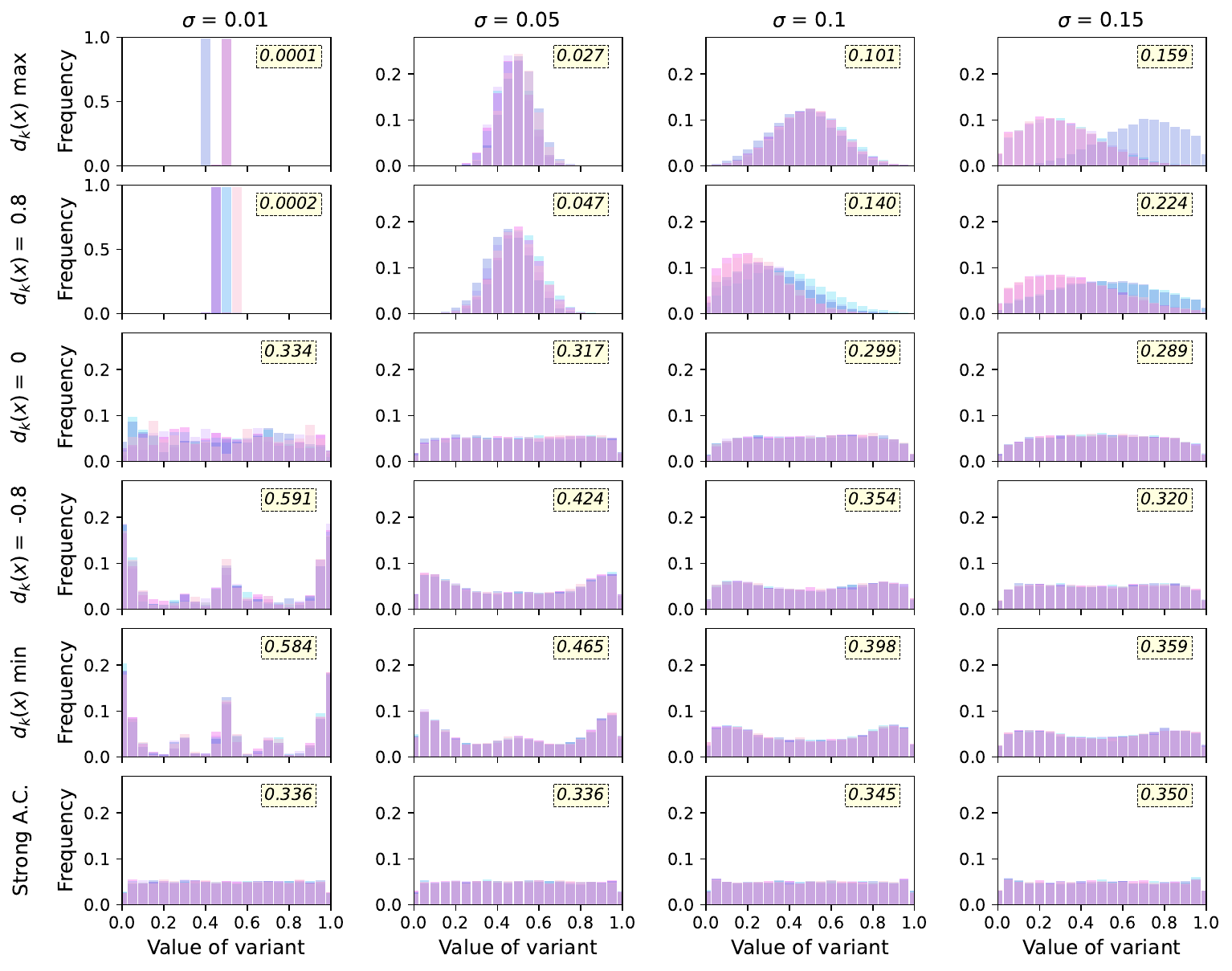}
  \caption{Similar to Figure \ref{linear.sigma.d.fixed.n}, except that the cultural trait is discrete rather than continuous. The trait can take values $0, 0.05, 0.1, 0.15, \dots, 1.0$. } \label{linear.sigma.d.fixed.n.discrete}
\end{figure*}

With conformity (top two rows of Figure \ref{linear.sigma.d.fixed.n}) and relatively low $\sigma$ values of $0.01$ and $0.05$, population distributions appear to peak around the center of the trait distribution. Notably, the mean cultural variant value in the initial population that was sampled from a uniform distribution on $[0,1]$ was 0.497 (Figure S1.1a). Thus, in these cases, conformity resulted in individuals becoming more similar to the population mean with lower population-level variance than the initial distribution. These dynamics are similar, but not identical, to previous models of conformity to a continuous trait in which conformity was defined as a preference for the mean variant in the population \cite{SmaldinoEpstein:2015, MorganThompson:2020}. 

However, none of the simulations in Figure \ref{linear.sigma.d.fixed.n} converged to a single point (which would be the case if every member of the population shared the same cultural variant), even when the conformity coefficient was very close to its upper bound (row 1 in  Figure \ref{linear.sigma.d.fixed.n}). This result differs from those in \cite{SmaldinoEpstein:2015}, where under conformity to a continuous trait (setting the ``distinctiveness preference'' parameter to zero for all individuals), the population converges to a single point. In our model of a \textit{discrete} cultural trait, however, a sufficiently high level of conformity combined with a sufficiently low standard deviation, $\sigma$, could produce convergence of nearly all members of the population to the same cultural trait value (e.g., Figure \ref{linear.sigma.d.fixed.n.discrete}, top left panel and the panel below it with bars for different replicates of the model near 1.0).

In both the continuous and discrete models, with conformity and sufficiently high $\sigma$, different replicates of the simulation may have peaks to the left or right of the center (e.g., far right, top two panels in Figure \ref{linear.sigma.d.fixed.n} and top right panel in Figure \ref{linear.sigma.d.fixed.n.discrete}), departing further from the expectations of models with population-mean transmission \cite{SmaldinoEpstein:2015, MorganThompson:2020}. In the continuous model where simulations ran for 1000 generations rather than 100, some of these distributions shifted, but not necessarily closer to the center of the plot (Figure S6.1).

\begin{figure*}[ht]
\raggedright
  \centering
  \includegraphics[width=12.5cm,height=9.4cm]{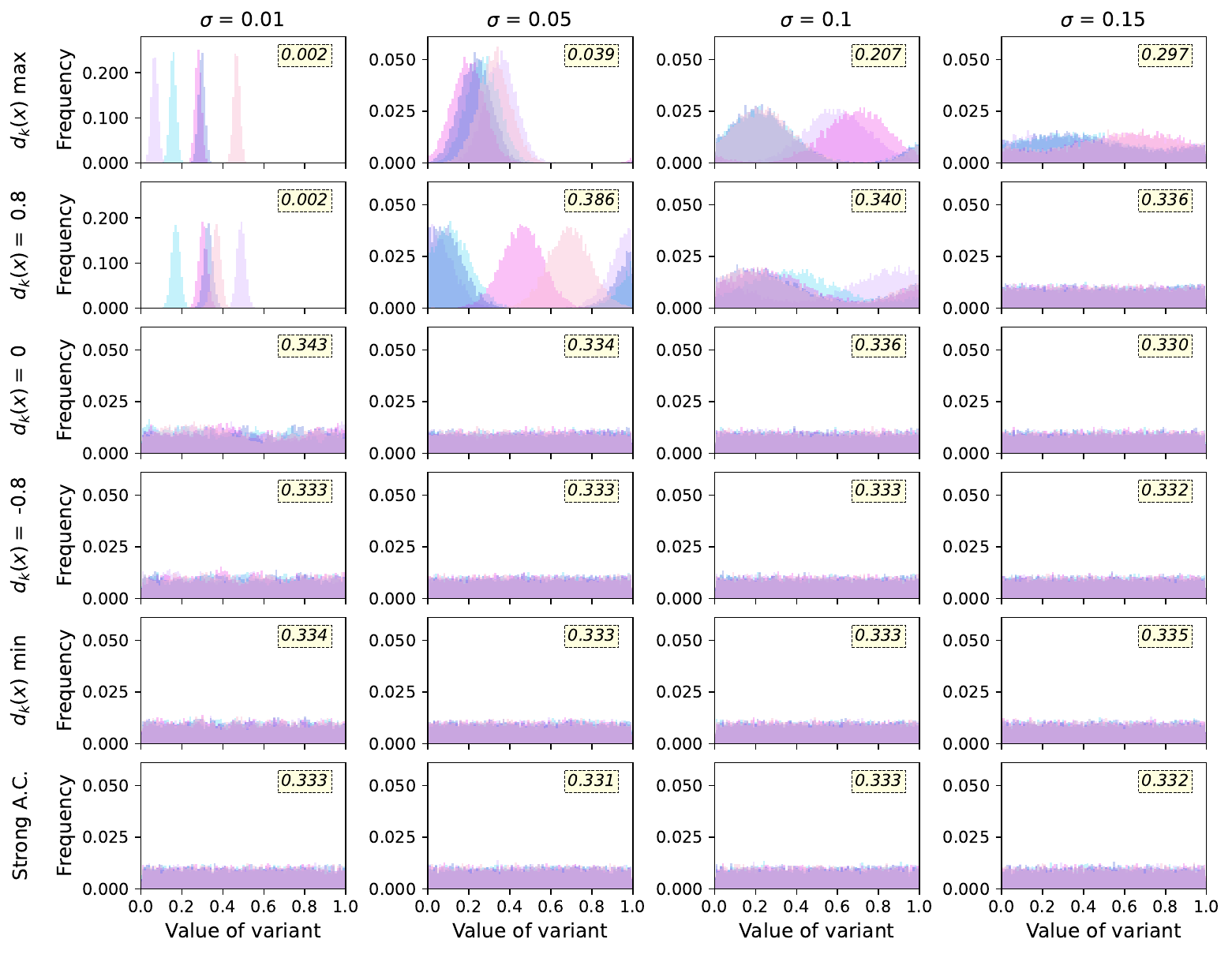}
  \caption{Similar to Figure \ref{linear.sigma.d.fixed.n}, except that the trait is circular rather than linear. } \label{circular.sigma.d.fixed.n}
\end{figure*}

What causes population distributions to remain at the left or right of the central point, 0.5, for many generations? One hypothesis is that ``edge effects'' play a role: in Figures \ref{linear.sigma.d.fixed.n}, \ref{linear.sigma.d.fixed.n.discrete}, S6.1, and S6.2, the cultural trait was linear (as in Figure \ref{linear.vs.circular}a) rather than circular. Consider what happens when an individual---either  conformist or anti-conformist---samples a role model whose variant $x_i$ is near one edge of the variant domain (0 or 1); say 0. First, the individual sets $x_i$ as the mean of a normal distribution with standard deviation $\sigma$. However, because the individual cannot sample a variant below 0, this normal distribution is re-normalized to $[0,1]$ (Definition 3), and therefore there is a disproportionately higher chance of sampling a variant near 0 than near other points. This causes the edges to exert a larger effect on the resulting distribution than the other points. In contrast, with a circular trait, samples from the normal distribution are not restricted to $[0,1]$ (see Section D, point 1), so there is no build-up of variants near the edges of the distribution after many generations (Figure \ref{circular.sigma.d.fixed.n}). This is especially notable in the case of anti-conformity; compare the left-most, second from the bottom panel of Figure \ref{linear.sigma.d.fixed.n} to that of Figure \ref{circular.sigma.d.fixed.n}. 

Some other differences between the linear model in Figure \ref{linear.sigma.d.fixed.n} and the circular model in Figure \ref{circular.sigma.d.fixed.n} are worth noting. For a circular trait, there is no ``mean'' value because there is no maximum or minimum value (Figure \ref{linear.vs.circular}b). Therefore, under conditions in which the population may move away from the boundaries and toward the mean in Figure \ref{linear.sigma.d.fixed.n}, this does not occur in Figure \ref{circular.sigma.d.fixed.n} (e.g., compare the top left panels of these two figures). In addition, the presence of \textit{two} boundaries (0 and 1), from which populations may move toward or away depending on the parameters, can produce symmetrical population distributions in Figure \ref{linear.sigma.d.fixed.n} that are absent from Figure \ref{circular.sigma.d.fixed.n} (e.g., compare row 2 from the top and column 2 from the left in both figures). 

In Figures \ref{linear.sigma.d.fixed.n}, \ref{linear.sigma.d.fixed.n.discrete}, S6.1, and S6.2, the model of strong anti-conformity (``Strong A.C.,'' bottom row) produces markedly different results from the model of anti-conformity defined in Eq. (\ref{eq.prob.xi}) (fourth and fifth rows). The key difference is that in the latter cases (fourth and fifth rows), anti-conformists prefer to adopt cultural variants that are near those of some other individuals (namely, individuals with variants that have high $k$-dispersals; Figure \ref{adopt_conformity_1}c,d), whereas strong anti-conformists prefer to adopt cultural variants that are far from those of all other individuals (Figure \ref{super_anticonform}). Thus, in a population of strong anti-conformists, ``clusters'' of individuals adopting similar cultural variants cannot be sustained, because each individual prefers to be as different as possible from the others. Thus, any peaks in the population distribution will likely flatten over time, and the resulting population will be near uniform.

\begin{figure*}[t]
\includegraphics[width=12.5cm,height=9.4cm]{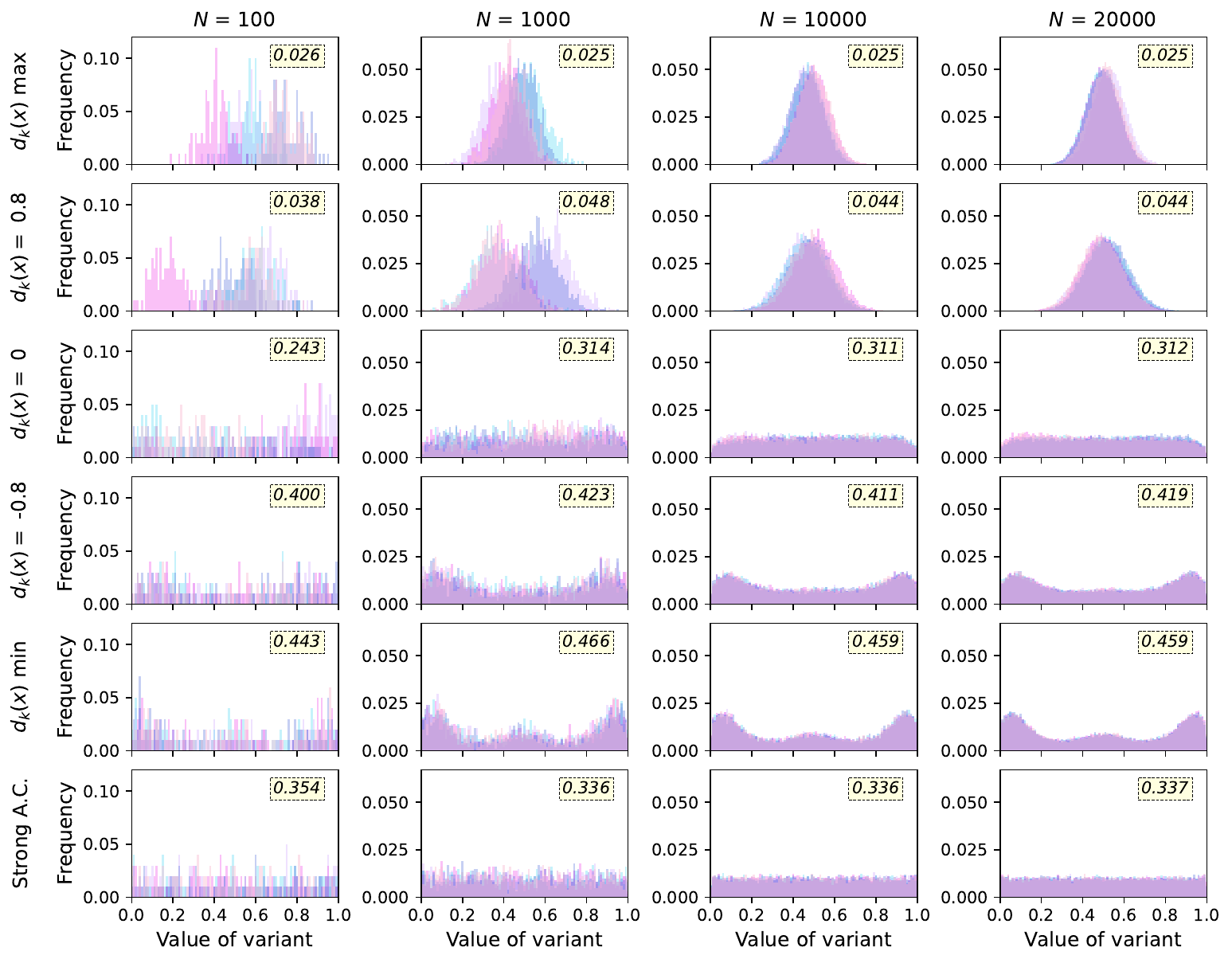}
  \caption{\small Effect of population size $N$ on population distributions after 100 generations. Here, the cultural trait is linear and continuous (or nearly continuous in the bottom row, where ``Strong A.C.'' stands for strong anti-conformity),  $k=2$, $n=5$, and $\sigma= 0.05$. In each panel, 5 replications were performed, each beginning from the same initial population that was generated from sampling $N$ individuals ($N$ is given in column labels) from a uniform distribution on $[0,1]$. Each of the 5 final distributions is plotted in a different color. For each of these 5 final distributions with cultural variants $\boldsymbol{x} = (x_1,x_2,\dots,x_N)$, the polarization index  $F = \frac{\text{Var}(\boldsymbol{x})}{\text{Mean}(\boldsymbol{x}) \cdot [1 - \text{Mean}(\boldsymbol{x})]}$ (see \cite{LeimarEtal:2022}) is calculated, and then the 5 $F$ values are averaged to produce the value shown in a yellow box in the top of each panel.  Values of $d_k(\boldsymbol{x})$ are in row labels. The label ``$d_k(\boldsymbol{x}) \min$'' refers to the case in which, for each $\boldsymbol{x}$, $d_k(\boldsymbol{x})$ is set to its lower bound (SI 2) plus $10^{-5}$. Similarly, ``$d_k(\boldsymbol{x}) \max$'' means that  $d_k(\boldsymbol{x})$ is set to its upper bound minus $10^{-5}$ (SI 2).   Histogram bin widths are 0.01, and $y$-axes differ in some of the panels (e.g., left-most column).
  } \label{linear.N.d}
\end{figure*}

Figure \ref{linear.N.d} shows the effects of varying the population size, $N$, and the conformity coefficients, $d_k(\boldsymbol{x})$, holding constant $k=2$, $n=5$, and $\sigma = 0.05$. Population distributions after 100 generations appear relatively similar with $N=1000$, $N=10,000$, and $N=20,000$. With $N=100$, distributions appear more `spiked' or less smooth than those with higher $N$ values, but nevertheless shapes and average polarity measures are often similar to the distributions with higher $N$. 

\begin{figure*}[t]
\centering
\includegraphics[width=12.5cm,height=9.4cm]{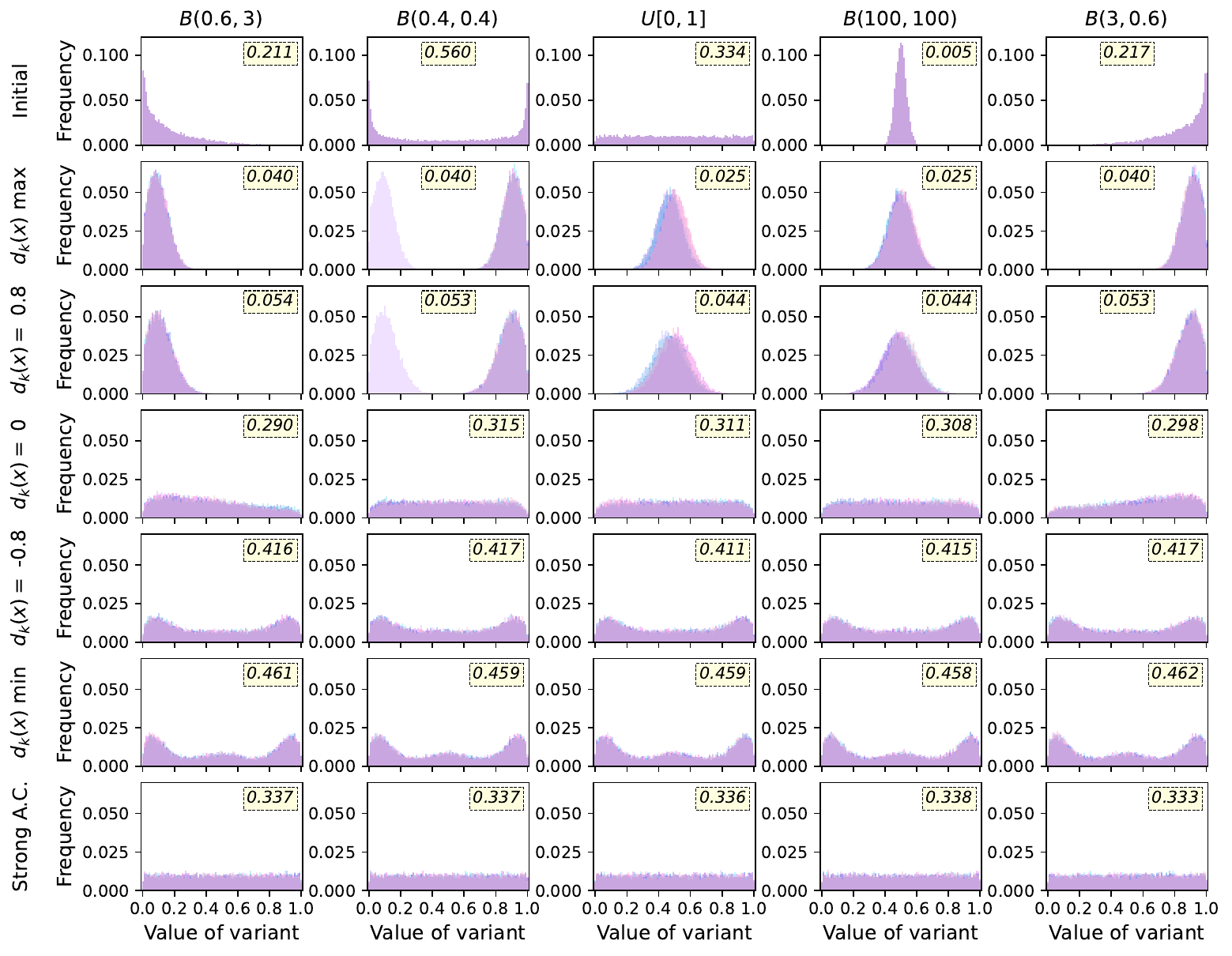}
  \caption{\small Effect of initial population distributions (top row) on distributions after 100 generations (rows 2-7) for a linear trait, with $k=2$, $n=5$, $N=10,000$, and $\sigma = 0.05$. In the top row, 5 initial population distributions are shown; these were sampled from the probability distributions in the column labels, where $B(\alpha, \beta)$ denotes a beta distribution with parameters $\alpha$ and $\beta$, and $U[0,1]$ denotes a uniform distribution on $[0,1]$. In rows 2 through 7, a panel in column $j$ shows 5 replications of the simulation, each beginning from the initial population shown in row 1, column $j$ and each in a different color. For each of these 5 final distributions with cultural variants $\boldsymbol{x} = (x_1,x_2,\dots,x_N)$, the polarization index  $F = \frac{\text{Var}(\boldsymbol{x})}{\text{Mean}(\boldsymbol{x}) \cdot [1 - \text{Mean}(\boldsymbol{x})]}$ (see \cite{LeimarEtal:2022}) is calculated, and these 5 $F$ values are averaged to produce the value shown in a yellow box in the top of each panel.  Values of $d_k(\boldsymbol{x})$ are in row labels. The label ``$d_k(\boldsymbol{x}) \min$'' refers to the case in which, for each $\boldsymbol{x}$, $d_k(\boldsymbol{x})$ is set to its lower bound (SI 2) plus $10^{-5}$. Similarly, ``$d_k(\boldsymbol{x}) \max$'' means that  $d_k(\boldsymbol{x})$ is set to its upper bound minus $10^{-5}$ (SI 2).   Histogram bin widths are 0.01, and $y$-axes differ in some of the panels (e.g., row 1).}\label{vary_d_initial}
\end{figure*}

Finally, in Figure \ref{vary_d_initial}, we hold constant $k=2$, $n=5$, $\sigma = 0.05$, and $N=10,000$, and vary the conformity coefficients $d_k(\boldsymbol{x})$ (rows) as well as the initial distribution of cultural variants in the population (columns). Each initial population distribution is shown in the top row. In the bottom three rows of Figure \ref{vary_d_initial}, with anti-conformity, population distributions after 100 generations are very similar across columns (i.e., regardless of different initial conditions). With random copying, population distributions after 100 generations differ slightly in cases where initial distributions are skewed to the left (first column) or to the right (last column). However, with conformity, population distributions after 100 generations often depend sensitively on the initial population composition. There is one exception; comparing cases in which the population is initially uniform vs. initially clustered in the center of the trait distribution (columns 3 and 4, respectively), there are similar trait distributions after 100 generations. This is not simply due to the fact that initial distributions in columns 3 and 4 have similar means (0.497 and 0.500, respectively) because the initial distribution in column 2 also has a similar mean of 0.501; rather, both the mean and density of the initial distribution affect evolutionary trajectories.  \vspace{-0.5cm}

\section{Discussion} \vspace{-0.2cm}

\noindent Research on frequency-dependent cultural transmission to date has been primarily focused on nominal  traits, with much less attention paid to ordinal traits, i.e., those with ordered variants. Here, we have explored three types of frequency-dependent cultural transmission, namely conformity, anti-conformity, and random copying, and ordinal traits that are continuous or discrete and linear or circular (Table 1). \textit{Linear} traits have minimum and maximum values, such as the proportion of resources shared with a friend ($0\%$ to $100\%$) or the amount of time spent on a task ($0$ seconds to one's entire lifetime), whereas circular traits do not; e.g., season of the year for an event or time on a clock chosen for sleeping. Even if arbitrary minima and maxima are set for a circular trait (e.g., 11:59 p.m. as the maximum and 12:00 a.m. as the minimum time on a clock), the fact that the minimum and maximum are \textit{more similar to each other} than they are to points in the middle of the distribution (say, noon) means that the trait can be conceptualized as a ``circle'' with adjoining ends rather than a line (Figure \ref{linear.vs.circular}).

At least three previous studies have considered conformity to an ordinal trait; specifically, a continuous, linear trait \cite{CavalliSforzaFeldman:1973, SmaldinoEpstein:2015, MorganThompson:2020} (note that \cite{CavalliSforzaFeldman:1973} did not use the term ``conformity,'' although their formalization is similar to \cite{SmaldinoEpstein:2015} and \cite{MorganThompson:2020}, which did). Morgan and Thompson \cite{MorganThompson:2020} define conformity as ``a tendency to adopt the mean trait
value in a population with an expected squared error less than the population variance (thereby causing the population to homogenize)'' (p. 2). They assume that all individuals know the variance of trait values in the entire population, but not the mean trait value in the population; instead, individuals estimate this mean by sampling a number of cultural role models (corresponding to $n$ in the present study). Morgan and Thompson show that under their definition of conformity, population-level variation decreases over time until all individuals in the population share the same cultural trait value. 

In the model of Smaldino and Epstein \cite{SmaldinoEpstein:2015}, an individual's ideal trait value is the mean trait value in the population, which they denote by $\bar{x}(t)$, plus the product of the standard deviation of trait values in the population, $\sigma(t)$, and the individual $i$'s ``distinctiveness preference,'' $\delta_i$. A non-zero distinctiveness preference corresponds to some extent of anti-conformity, i.e., preferring to be different from the mean trait in the population, whereas ``conformists prefer to be at the population mean'' with $\delta_i = 0$ (\cite{SmaldinoEpstein:2015}, p. 6). Unlike Morgan and Thompson \cite{MorganThompson:2020}, all individuals in the model of Smaldino and Epstein \cite{SmaldinoEpstein:2015} know both the mean trait value in the population, $\bar{x}(t)$, and the standard deviation of trait values in the population, $\sigma(t)$. They present several models that differ in the distribution of distinctiveness preferences, $\delta_i$, across individuals, summarized in their Table 1. Among other findings, they show that if all members of the population have the same distinctiveness preference ($\delta_i = \delta \ \forall \ i$), then at equilibrium all individuals share the same cultural trait value. Therefore, if all individuals are entirely conformist ($\delta_i = 0$), then, as in \cite{MorganThompson:2020}  there is no variance in the trait distribution at equilibrium.  \vspace{-0.1cm}

However, the definition of ``conformity'' for continuous traits proposed by \cite{SmaldinoEpstein:2015} and \cite{MorganThompson:2020}, namely population-mean transmission, is not analogous to the definition of conformity for nominal traits proposed by Boyd and Richerson \cite{BoydRicherson:1985} and used extensively throughout the conformity literature (e.g., \cite{HenrichBoyd:1998, Henrich:2001, HenrichBoyd:2001, KamedaNakanishi:2002, Nakahashi:2007, WakanoAoki:2007, MollemanEtal:2013, BorofskyFeldman:2022, DentonEtal:2020, DentonEtal:2021, LappoEtal:2023, DentonEtal:2021b, DentonEtal:2023}). Boyd and Richerson define conformity as a disproportionate tendency to copy the majority; e.g., if 60\% of individuals in the population are performing a behavior, a conformist adopts this behavior with a probability greater than 60\% \cite{BoydRicherson:1985}. To illustrate how large the difference between this definition of conformity and conformity as population-mean transmission can be, consider the cultural trait ``proportion of resources shared with others.'' Assume that 60\% of the population shares \textit{all} of their resources, whereas the other 40\% of the population shares \textit{none}. Under Boyd and Richerson's definition, conformists would either choose to share \textit{all} or \textit{none} of their resources (i.e., the behaviors of others they have seen), and would have a greater-than-60\% chance of sharing \textit{all} their resources. On the other hand, if conformity were conceptualized as population-mean transmission, then conformists would be expected to share a fraction of $(60\% \times 100\%) + (40\% \times 0\%) = 60\%$ of their resources---a behavior that they have not seen. Our goal was to develop a framework for studying conformity to ordinal traits (including linear and continuous traits as in  \cite{SmaldinoEpstein:2015, MorganThompson:2020}) that did not rely on population-mean transmission, but instead captured the idea that conformists preferentially adopt more popular cultural traits. \vspace{-0.1cm}

Determining the ``commonness'' of a cultural variant by counting the number of individuals that share that cultural variant may not be useful for continuous traits with infinitely many variants, because it is highly likely that all individuals have different variants. Therefore, instead of commonness, we assess how clustered cultural variants are in phenotypic space by introducing a metric called $k$-dispersal (the sum of the $k$ shortest absolute distances from a given cultural variant to the other observed cultural variants). Conformity can then be defined as a disproportionate tendency to adopt less dispersed (more clustered) cultural variants and anti-conformity as the opposite tendency (Figure \ref{adopt_conformity_1}). In this case, unlike in the models of  \cite{SmaldinoEpstein:2015} and \cite{MorganThompson:2020}, conformity does not lead all members of the population to converge on the same cultural variant. Instead, populations continue for hundreds of generations to display a range of different cultural variants (Figures \ref{linear.sigma.d.fixed.n} and S6.1), which may be narrow (as in the top left of Figures \ref{linear.sigma.d.fixed.n} and S6.1) or broad (as in the top right of Figures \ref{linear.sigma.d.fixed.n} and S6.1). These distributions may be centered around a value that is close to the mean trait value in the initial population (e.g., the initial mean was 0.497 in Figures \ref{linear.sigma.d.fixed.n} and S6.1, similar to the top left panel) or around a different value (e.g., top right of Figures \ref{linear.sigma.d.fixed.n} and S6.1). 

Similarly, with discrete rather than continuous linear traits, population distributions after many generations may be either broad (top right of Figure \ref{linear.sigma.d.fixed.n.discrete}) or narrow (top left of Figure \ref{linear.sigma.d.fixed.n.discrete}). In contrast to continuous traits for which there are infinitely many variants, for the discrete trait shown in Figure \ref{linear.sigma.d.fixed.n.discrete} there are only 21 possible variants: $0, \tfrac{1}{20}, \tfrac{2}{20}, \tfrac{3}{20}, \dots, 1$, and it is possible that multiple individuals share the same cultural variant (in fact, in a population with $N > 21$ individuals, it is guaranteed). In the discrete model with conformist transmission ($d_k(\boldsymbol{x})>0$) and high precision in trait adoption (low $\sigma$ in Definition 3), populations can reach a point where nearly all individuals share the same cultural variant. Compared to our continuous trait model, these results from the discrete model align more closely with the expectations of \cite{SmaldinoEpstein:2015} and \cite{MorganThompson:2020}. Similarly, in models of nominal traits, conformist transmission often results in all individuals eventually sharing the same cultural variant (some exceptions are discussed in \cite{DentonEtal:2022}). 

In addition to the definition of conformity and anti-conformity, another difference between our models and those of Smaldino and Epstein \cite{SmaldinoEpstein:2015} is that we do not assume that individuals know the mean and standard deviation of trait values in the whole population. In the real world, these values may be difficult to estimate as the number $N$ of individuals in the population becomes large. Instead, following \cite{BoydRicherson:1985} (see also \cite{MorganThompson:2020}), we assume that each individual takes a random sample of $n < N$ individuals and uses this local information to estimate what is ``popular'' and ``unpopular'' in the population. Therefore, it is possible that some individuals have an inaccurate sense of traits' popularities. For instance, a conformist might, by chance, sample $n=6$ models as in Figure \ref{adopt_conformity_1}b and believe that cultural variants near 0.1-0.15 are quite popular, even if the wider population has very few cultural variants in this range. 

When an individual chooses a cultural variant from a sample of $n$ role models as in Figures \ref{variant_adopt} and \ref{adopt_conformity_1}, the precision with which a value near one of the role models' values is chosen is governed by the parameter $\sigma$ (see Definitions 1 and 3). These figures illustrate the potential cultural variant choices, and their probabilities, for one individual in one generation, but do not depict the evolutionary path over multiple generations and multiple individuals. Smaller values of $\sigma$ produce tighter peaks around the sampled variants (right-hand sides of Figures \ref{variant_adopt} and \ref{adopt_conformity_1}) whereas larger values of $\sigma$ produce wider peaks and a smoother, less spiky distribution (left-hand sides of Figures \ref{variant_adopt} and \ref{adopt_conformity_1}). A larger standard deviation allows for greater flexibility and exploration in the adoption of cultural traits, leading to a broader spectrum of potential trait selections. Therefore, in the multi-individual and multi-generation evolutionary simulations, higher $\sigma$ values produce smoother population distributions, with lower peaks and higher valleys, compared to lower $\sigma$ values (Figures \ref{linear.sigma.d.fixed.n}, \ref{linear.sigma.d.fixed.n.discrete}, \ref{circular.sigma.d.fixed.n}, S6.1, and S6.2). 

In our linear trait model with minimum and maximum values (0 and 1, respectively), anti-conformity according to Definition 3 may produce a significant build-up of cultural variants near the boundaries, especially when $\sigma$ is low (e.g., $\sigma = 0.01$ and $d_k(\boldsymbol{x}) < 0$ in the fourth and fifth rows of Figure \ref{linear.sigma.d.fixed.n}). Under anti-conformity this occurs for many different initial population distributions (e.g., Figure \ref{vary_d_initial}, fifth and sixth rows from the top). However, in the circular trait model in which values below 0 or above 1 are ``wrapped around'' to the opposite end of the distribution, this build-up does not occur, and anti-conformity produces a relatively uniform distribution across the trait space (Figure \ref{circular.sigma.d.fixed.n}). Similarly, when anti-conformity is represented not as in Definition 3 but instead as strong anti-conformity (e.g., Figure \ref{super_anticonform} and SI 3), build-up of density near the boundaries does not occur in either the linear or circular model (bottom row of Figures \ref{linear.sigma.d.fixed.n} and \ref{circular.sigma.d.fixed.n}). This is likely due to the fact that in Definition 3, anti-conformists prefer to be similar to \textit{some} individuals (namely, those whose cultural variants are less clustered in the trait space; see the bottom row of Figure \ref{adopt_conformity_1}) whereas strong anti-conformists do not prefer to be similar to \textit{any} individuals, so build-ups in the frequencies of cultural variants are not sustained over time with strong anti-conformity.  

Our extension of previous conceptions of conformity and anti-conformity with nominal traits---namely tendencies to adopt more or less popular traits, respectively---to linear and circular ordinal traits that can be discrete or continuous has produced an interesting array of evolutionary outcomes. Future research could expand on these models in several ways. For example, our models included a single population of size $N$, whereas in reality there may be different sub-groups, each with a different level of conformity, interacting as a meta-population. At a finer scale, different individuals within sub-groups could vary in their tendencies to conform or anti-conform, in which case $d_k(\boldsymbol{x})$ could be chosen from a probability distribution. Exploring alternative approaches for the role model sampling process is another natural extension of our model. Rather than sampling $n$ role models at random, an individual could select role models based on  prestige, success, or similarity to  themselves. 

\vspace{0.1cm} \text{  }

\section*{Acknowledgments} \vspace{-0.2cm}

\noindent This research was supported in part by the Morrison Institute for Population and Research Studies at Stanford University; the Stanford Center for Computational, Evolutionary and Human Genomics; and the Santa Fe Institute. We acknowledge participants at the Modeling and Theory in Population Biology meeting of the Banff International Research Station (24htp001) for helpful suggestions.


\end{document}